\documentclass[aps,pra,reprint,preprintnumbers,amsmath,amssymb,superscriptaddress,showpacs,showkeys,longbibliography]{revtex4-1}

\usepackage{graphicx}
\usepackage{dcolumn}
\usepackage{bm}
\usepackage{color}
\usepackage{braket}
\usepackage{units}
\usepackage[english]{babel}
\usepackage[
	colorlinks=true,
	frenchlinks=false,
        linkcolor=blue,
	    anchorcolor=blue,
        citecolor=blue,
        filecolor=blue,
        urlcolor=blue,
        bookmarks=true,
        bookmarksopen=true,
	    bookmarksnumbered=true, 
        bookmarksopenlevel=1,
        plainpages=false,
        pdfpagelabels=true,
	    draft=false]{hyperref}

\begin{document}

\title[]{Flux-driven Josephson Parametric Amplifiers: Hysteretic Flux Response and Nondegenerate Gain Measurements}

\author{\firstname{Stefan}~\surname{Pogorzalek}}
\email[]{stefan.pogorzalek@wmi.badw.de}
\affiliation{Walther-Mei{\ss}ner-Institut, Bayerische Akademie der Wissenschaften, 85748 Garching, Germany }
\affiliation{Physik-Department, Technische Universit\"{a}t M\"{u}nchen, 85748 Garching, Germany}
\author{Kirill G. Fedorov}
\affiliation{Walther-Mei{\ss}ner-Institut, Bayerische Akademie der Wissenschaften, 85748 Garching, Germany }
\affiliation{Physik-Department, Technische Universit\"{a}t M\"{u}nchen, 85748 Garching, Germany}
\author{Ling Zhong}
\affiliation{Walther-Mei{\ss}ner-Institut, Bayerische Akademie der Wissenschaften, 85748 Garching, Germany }
\affiliation{Physik-Department, Technische Universit\"{a}t M\"{u}nchen, 85748 Garching, Germany}
\affiliation{Nanosystems Initiative Munich (NIM), Schellingstra{\ss}e 4, 80799 M\"{u}nchen, Germany}
\author{Jan Goetz}
\affiliation{Walther-Mei{\ss}ner-Institut, Bayerische Akademie der Wissenschaften, 85748 Garching, Germany }
\affiliation{Physik-Department, Technische Universit\"{a}t M\"{u}nchen, 85748 Garching, Germany}
\author{Friedrich Wulschner}
\affiliation{Walther-Mei{\ss}ner-Institut, Bayerische Akademie der Wissenschaften, 85748 Garching, Germany }
\affiliation{Physik-Department, Technische Universit\"{a}t M\"{u}nchen, 85748 Garching, Germany}
\author{Michael Fischer}
\affiliation{Walther-Mei{\ss}ner-Institut, Bayerische Akademie der Wissenschaften, 85748 Garching, Germany }
\affiliation{Physik-Department, Technische Universit\"{a}t M\"{u}nchen, 85748 Garching, Germany}
\affiliation{Nanosystems Initiative Munich (NIM), Schellingstra{\ss}e 4, 80799 M\"{u}nchen, Germany}
\author{Peter Eder}
\affiliation{Walther-Mei{\ss}ner-Institut, Bayerische Akademie der Wissenschaften, 85748 Garching, Germany }
\affiliation{Physik-Department, Technische Universit\"{a}t M\"{u}nchen, 85748 Garching, Germany}
\affiliation{Nanosystems Initiative Munich (NIM), Schellingstra{\ss}e 4, 80799 M\"{u}nchen, Germany}
\author{Edwar Xie}
\affiliation{Walther-Mei{\ss}ner-Institut, Bayerische Akademie der Wissenschaften, 85748 Garching, Germany }
\affiliation{Physik-Department, Technische Universit\"{a}t M\"{u}nchen, 85748 Garching, Germany}
\affiliation{Nanosystems Initiative Munich (NIM), Schellingstra{\ss}e 4, 80799 M\"{u}nchen, Germany}
\author{Kunihiro Inomata}
\affiliation{RIKEN Center for Emergent Matter Science (CEMS), Wako, Saitama 351-0198, Japan}
\author{Tsuyoshi Yamamoto}
\affiliation{NEC IoT Device Research Laboratories, Tsukuba, Ibaraki 305-8501, Japan}
\author{Yasunobu Nakamura}
\affiliation{Research Center for Advanced Science and Technology (RCAST), The University of Tokyo, Meguro-ku, Tokyo 153-8904, Japan}
\author{Achim Marx}
\affiliation{Walther-Mei{\ss}ner-Institut, Bayerische Akademie der Wissenschaften, 85748 Garching, Germany }
\author{Frank Deppe}
\affiliation{Walther-Mei{\ss}ner-Institut, Bayerische Akademie der Wissenschaften, 85748 Garching, Germany }
\affiliation{Physik-Department, Technische Universit\"{a}t M\"{u}nchen, 85748 Garching, Germany}
\affiliation{Nanosystems Initiative Munich (NIM), Schellingstra{\ss}e 4, 80799 M\"{u}nchen, Germany}
\author{Rudolf Gross}
\email[]{rudolf.gross@wmi.badw.de}
\affiliation{Walther-Mei{\ss}ner-Institut, Bayerische Akademie der Wissenschaften, 85748 Garching, Germany }
\affiliation{Physik-Department, Technische Universit\"{a}t M\"{u}nchen, 85748 Garching, Germany}
\affiliation{Nanosystems Initiative Munich (NIM), Schellingstra{\ss}e 4, 80799 M\"{u}nchen, Germany}

\date{\today}

\begin{abstract}
Josephson parametric amplifiers (JPA) have become key devices in quantum science and technology with superconducting circuits. In particular, they can be utilized as quantum-limited amplifiers or as a source of squeezed microwave fields. Here, we report on the detailed measurements of five flux-driven JPAs, three of them exhibiting a hysteretic dependence of the resonant frequency versus the applied magnetic flux. We model the measured characteristics by numerical simulations based on the two-dimensional potential landscape of the dc superconducting quantum interference devices (dc-SQUID), which provide the JPA nonlinearity, for a finite screening parameter $\beta_\mathrm{L}\,{>}\,0$ and demonstrate excellent agreement between the numerical results and the experimental data. Furthermore, we study the nondegenerate response of different JPAs and accurately describe the experimental results with our theory.
\end{abstract}



\maketitle
\section{Introduction}
The field of superconducting quantum circuits (SQC) is a highly active field of research with outstanding progress over the last decade. Typically, the circuits are operated at microwave frequencies and therefore profit from commercially available low-noise amplifiers. However, the operation of SQCs often requires quantum-limited amplification of weak signals over a broad frequency range. For this purpose, one can either use broadband traveling wave parametric amplifiers~\cite{Macklin:2015,white:2015} or band tunable narrow-band JPAs~\cite{Yamamoto:2008,mutus:2014,eichler:2014,Roch:2012}.
The latter are nowadays routinely used to provide amplification with a noise performance close to the standard quantum limit~\cite{Yurke:1996,Castellanos-Beltran:2007,Bergeal:2010,Hatridge:2011,Zhong:2013,Simoen:2015}. 
Moreover, if operated in the phase sensitive regime, JPAs can even achieve noise temperatures below the standard quantum limit~\cite{Yurke:1989a,Zhong:2013,Castellanos-Beltran:2008}.
JPAs also have been shown to generate squeezed microwave light~\cite{Menzel:2010,Eichler:2011b,Mallet:2011,Tholen:2009,Yurke:2006,Menzel:2012}, which can be be utilized in numerous quantum information processing algorithms. A particularly prominent application is the generation of entanglement in the form of two-mode squeezed propagating microwave states~\cite{Menzel:2012,Fedorov:2016,Flurin:2012} as required for many quantum communication and quantum teleportation protocols with continuous variables~\cite{Candia:2015}. Also, the bipartite entanglement of the emitted light modes allows one to detect low reflectivity objects in quantum illumination protocols more efficiently than in the classical case~\cite{Lloyd:2008,Tan:2008}. Furthermore, squeezed states are of importance in the context of quantum computing with continuous variables~\cite{Fedorov:2016,Andersen:2015,Weedbrook:2012}.

In this work, we study JPAs consisting of a superconducting coplanar waveguide resonator short-circuited to ground by a dc-SQUID. In particular, we investigate how the resonant frequency of five JPAs with different screening parameters of the constituent dc-SQUID depends on the magnetic flux threading the dc-SQUID loop. From numerical simulations of the dc-SQUID potential, we find that the flux dependence of the JPA resonant frequency is, in general, hysteretic for arbitrary screening parameters of the dc-SQUID and find good agreement with experimental observations. Furthermore, we report on the nondegenerate gain properties when applying a pump tone to the JPAs. The JPA response depends strongly on the resonator characteristics and coincides accurately with our theoretical predictions for a flux-driven JPA.

The rest of this paper is structured as follows. An introduction of the flux-driven JPA in Sec.~\ref{sec:JPAGeneral} is followed by a theoretical description of the dependence of the JPA resonant frequency on an external flux in Sec.~\ref{sec:theory JPA}. The behavior of the dc-SQUID potential as a function of the external flux is described in Sec.~\ref{sec:dcSQUIDPot}. In Sec.~\ref{sec:ExpResults}, we present the experimentally measured flux dependence of the JPA resonant frequency and compare it to theory. Finally, we investigate the nondegenerate gain properties for two JPAs with different parameters in Sec.~\ref{sec:GainMeasurements} and summarize our results in Sec.~\ref{sec:Summary}.

\section{The flux-driven JPA}\label{sec:JPAGeneral}

\begin{figure}
        \begin{center}
        \includegraphics{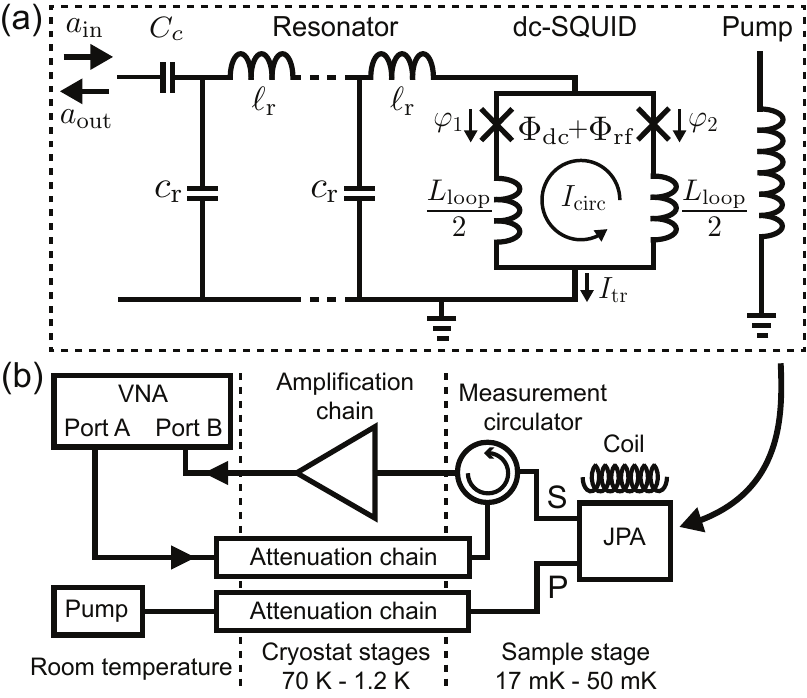}
        \end{center}
    \caption{(a) Circuit diagram of a JPA consisting of a coplanar waveguide resonator with inductance $\ell_\mathrm{r}$ and capacitance $c_\mathrm{r}$ per unit length which add up to the total inductance $L_\mathrm{r}$ and capacitance $C_\mathrm{r}$. The dc-SQUID loop has a finite geometrical loop inductance $L_\mathrm{loop}$ and, thus, inductively couples to a magnetic flux $\Phi_\mathrm{dc}\,{+}\,\Phi_\mathrm{rf}$ generated by a coil and a pump line, respectively. Crosses indicate Josephson junctions. (b) Setup for the characterization of JPAs with a VNA. The reflected signal from the JPA is separated from the input signal by a measurement circulator. The current through a superconducting coil determines the flux $\Phi_\mathrm{dc}$ through the dc-SQUID loop. S and P mark the signal and pump port, respectively.}
    \label{fig1}
\end{figure}

We investigate a flux-driven JPA~\cite{Yamamoto:2008} consisting of a quarter-wavelength coplanar waveguide resonator which is short-circuited to ground by a dc-SQUID [see Fig.~\ref{fig1}(a)]. The dc-SQUID provides a flux-tunable non-linear inductance which contributes to the  quasi-static resonant frequency $\omega_0$ of the JPA. Thus, a magnetic flux can be used to tune the dc-SQUID inductance and, in this way, the resonant frequency of the whole circuit. An on-chip antenna couples inductively to the dc-SQUID loop with an inductance $L_\mathrm{loop}$ and is used to apply a strong coherent pump tone with a frequency $\omega_\mathrm{pump}\,{=}\,2\omega_\mathrm{0}$.
The pump tone leads to a periodic modulation of the dc-SQUID inductance and, thus, to a periodic variation of the resonant frequency of the JPA around $\omega_0$ which gives rise to parametric effects~\cite{Louisell:1961,Gordon:1963}. A signal mode with frequency $\omega_\mathrm{s}\,{=}\,\omega_\mathrm{pump}/2\,{+}\,\delta \omega$ incident to the JPA input port is then amplified, while at the same time an idler mode at frequency ${\omega_\mathrm{i}\,{=}\,\omega_\mathrm{pump}/2-\delta\omega}$ is generated so that the total energy is conserved, $\omega_\mathrm{pump}\,{=}\,\omega_\mathrm{s}\,{+}\,\omega_\mathrm{i}$~\cite{Clerk:2010}.

In order to experimentally characterize the JPAs, we use the setup shown in Fig.~\ref{fig1}(b). A vector network analyzer~(VNA) applies a probe signal to the JPA via a measurement circulator, which separates the incoming signal from the signal reflected at the JPA. Subsequently, the reflected signal is amplified with cryogenic and room temperature amplifiers, and measured by the VNA to obtain the complex reflection coefficient $S_{11}$. The external magnetic field is generated by a superconducting coil. For nondegenerate gain measurements, a strong coherent pump tone is applied in addition to the VNA probe tone. 

\begin{figure}
        \begin{center}
        \includegraphics{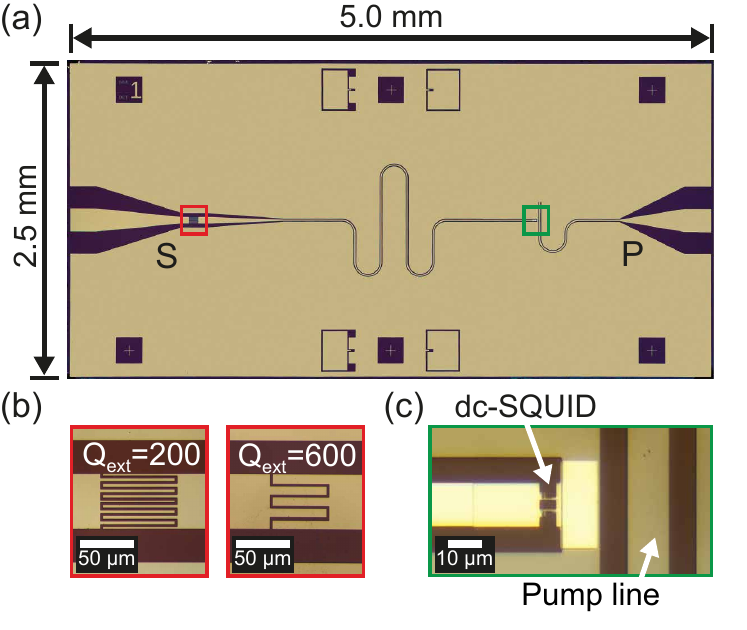}
        \end{center}
    \caption{(a) Optical micrograph of a JPA sample chip. Red and green rectangles mark the coupling capacitor and the dc-SQUID with the pump line, respectively. S and P mark the signal and pump ports, respectively. (b) Zoom-ins to coupling capacitors with a designed external quality factor of $Q_\mathrm{ext}\,{=}\,200$ (left), $Q_\mathrm{ext}\,{=}\,600$ (right). (c) Zoom-in to the dc-SQUID with the adjacent pump line. The size of the dc-SQUID loop is $4.2\times\unit[2.4]{\mu m}^2$.}
    \label{fig2}
\end{figure}

The JPA sample is encased inside a gold-plated copper box and mounted to the base temperature stage of a dilution refrigerator. In Fig.~\ref{fig2}, we show an optical micrograph of a JPA sample and magnified views of the coupling capacitor and the dc-SQUID region. All presented samples were fabricated at NEC and RIKEN, Japan. Resonator and pump line consist of a \unit[50]{nm} thick niobium film which is sputter-deposited onto a $\unit[300]{\mu m}$ thick thermally oxidized silicon substrate. Both resonator and pump line have a coplanar waveguide geometry with an impedance of $\unit[50]{\Omega}$. The dc-SQUID is fabricated with aluminum technology and double angle shadow evaporation~\cite{Dolan:1977}.

\section{Theory of the resonant frequency of a tunable JPA}\label{sec:theory JPA}
In this section, we describe how the resonant frequency of the JPA circuit depends on an external magnetic flux $\Phi_\mathrm{ext}$ threading the dc-SQUID loop. The treatment is applicable for arbitrary flux-screening of the dc-SQUID, which can be quantified by the dimensionless screening parameter $\beta_\mathrm{L}$ defined below. Based on a distributed element model for a quarter-wavelength resonator and a lumped element model for the dc-SQUID [see Fig.~\ref{fig1}(a)], one arrives at a transcendental equation for the resonant frequency $\omega_0$ of the JPA~\cite{Wallquist:2006,Wustmann:2013}
\begin{equation}\label{eqn:WallquistW}
\frac{\pi \omega_0}{2\omega_\mathrm{r}}\tan\biggl(\!\frac{\pi \omega_0}{2\omega_\mathrm{r}}\!\biggr)\,{=}\,\frac{(2\pi)^2}{\Phi_0^2}L_\mathrm{r}E_s(\Phi_\mathrm{ext})\,{-}\,\frac{2C_\mathrm{s}}{C_\mathrm{r}}\biggl(\!\frac{\pi \omega_0}{2\omega_\mathrm{r}}\!\biggr)^{\!2} ,
\end{equation}
where $\omega_\mathrm{r}$, $L_\mathrm{r}$ and $C_\mathrm{r}$ are the resonant frequency, the total inductance and capacitance of the bare resonator, respectively, $E_\mathrm{s}(\Phi_\mathrm{ext})$ is the flux-dependent energy of the dc-SQUID defined below, and $C_s$ is the capacitance of one Josephson junction. For the investigated samples, the last term in Eq.~(\ref{eqn:WallquistW}) can be neglected, since the capacitance of the resonator strongly exceeds the one of the Josephson junctions, $C_\mathrm{r}\gg C_\mathrm{s}$. For a vanishing transport current $I_\mathrm{tr}$ through the dc-SQUID, we define the Josephson inductance of the dc-SQUID as
\begin{equation}
L_\mathrm{s}(\Phi_\mathrm{ext})=\frac{\Phi_0}{4 \pi I_\mathrm{c}|\cos\varphi_-^\mathrm{min}(\Phi_\mathrm{ext})|}\, ,
\end{equation}
where $I_\mathrm{c}$ is the critical current of a single Josephson junction and $\varphi_\pm\,{\equiv}\,(\varphi_1\pm\varphi_2)/2$ is derived from the phase differences $\varphi_1$ and $\varphi_2$ across each of the Josephson junctions of the dc-SQUID. Here, $\varphi_\pm^\mathrm{min}(\Phi_\mathrm{ext})$ are the steady-state phase differences for a given external flux $\Phi_\mathrm{ext}$.
Then, one can express the flux-dependent energy of the dc-SQUID as
\begin{equation}\label{eqn:energyEs}
E_\mathrm{s}(\Phi_\mathrm{ext})=\frac{\Phi_0^2}{(2\pi)^2}\frac{1}{L_\mathrm{s}(\Phi_\mathrm{ext})+L_\mathrm{loop}/4} \, ,
\end{equation}
where also the finite dc-SQUID loop inductance $L_\mathrm{loop}$ is taken into account~\cite{Bhupathi:2016}.
 
The tangent in Eq.~(\ref{eqn:WallquistW}) can be expanded around $\omega_0/\omega_\mathrm{r}\simeq 1$ to obtain a simplified expression for the resonant frequency of the JPA in terms of inductances
\begin{equation}\label{eqn:WallquistWApprox}
\omega_0(\Phi_\mathrm{ext})=\omega_\mathrm{r}\left[1+\cfrac{L_\mathrm{s}(\Phi_\mathrm{ext})+L_\mathrm{loop}/4}{L_\mathrm{r}}\right]^{-1}\, .
\end{equation}
We now discuss the effect of flux-screening of the dc-SQUID. In general, $\varphi_-^\mathrm{min}(\Phi_\mathrm{ext})$ exhibits a non-trivial dependence on the external magnetic flux. Due to the fluxoid quantization, $\varphi_-^\mathrm{min}(\Phi_\mathrm{ext})$ is related to the total flux $\Phi\,{=}\,\Phi_\mathrm{ext}\,{+}\,L_\mathrm{loop}I_\mathrm{circ}$ threading the dc-SQUID as $\varphi_-^\mathrm{min}\,{=}\,\pi(\Phi/\Phi_0)$. The total flux is given by the sum of the external flux $\Phi_\mathrm{ext}$ and a part $\Phi_\mathrm{circ}\,{=}\,L_\mathrm{loop}I_\mathrm{circ}$ originating from the supercurrent $I_\mathrm{circ}$ circulating in the dc-SQUID loop. To account for this screening current, we introduce the dimensionless screening parameter
\begin{equation}
\beta_\mathrm{L}\equiv\frac{2L_\mathrm{loop}I_\mathrm{c}}{\Phi_0}\, .
\end{equation} 

In the case of a vanishing screening parameter $\beta_\mathrm{L}\,{\simeq}\,0$, we have $\Phi\,{\approx}\,\Phi_\mathrm{ext}$. In this case, the fluxoid quantization fixes the phase of one Josephson junction relative to the other one, reducing the available degrees of freedom from two to one. This results in a single-valued dependence $\varphi_-^\mathrm{min}\,{=}\,\pi(\Phi_\mathrm{ext}/\Phi_0)$ and one obtains the well-known expression for the flux-dependent inductance of a dc-SQUID, $L_\mathrm{s}(\Phi_\mathrm{ext})\,{=}\,\Phi_0/(4 \pi I_\mathrm{c}|\cos\pi(\Phi_\mathrm{ext}/\Phi_0)|)$~\cite{Sandberg:2008}. A mechanical analog of this situation are two rigidly coupled pendula, where due to the rigid coupling the system can be described by a single deflection angle, i.e., a single degree of freedom. 
However, if the screening parameter $\beta_\mathrm{L}$ becomes finite, there is no analytic expression for $\varphi_-^\mathrm{min}(\Phi_\mathrm{ext})$ anymore and the dependence has to be calculated numerically.

\section{Simulation of the dc-SQUID potential}\label{sec:dcSQUIDPot}

\begin{figure}
        \begin{center}
        \includegraphics{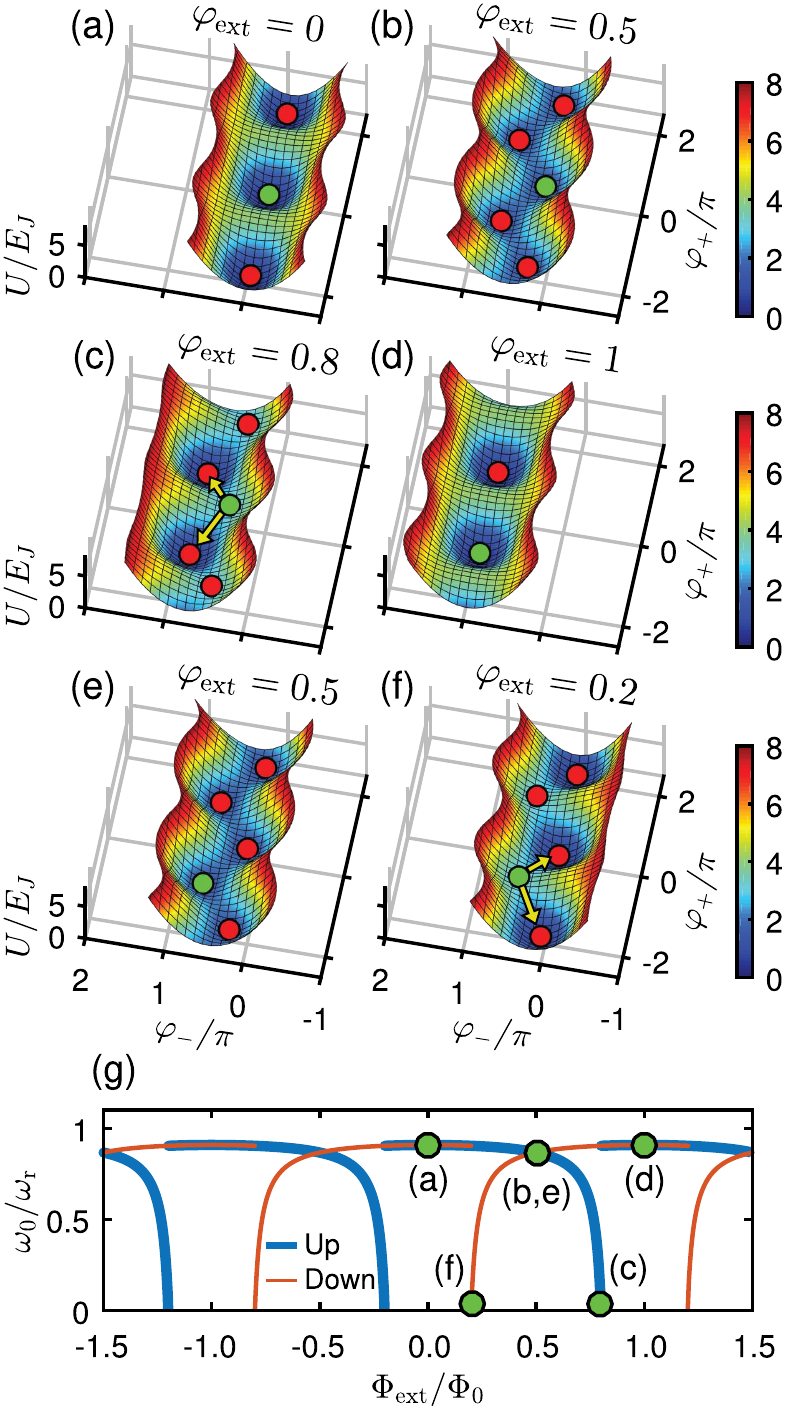}
        \end{center}
    \caption{(a-f) Two-dimensional dc-SQUID potential $U$ for different external flux values $\Phi_\mathrm{ext}$ with $\beta_\mathrm{L}\,{=}\,0.6$ and $j_\mathrm{tr}\,{=}\,0$. Green dots denote the present position of the phase particle, whereas red dots indicate the positions of all other local minima. Panels (a-d) and panels (e,f) correspond to increasing and decreasing $\Phi_\mathrm{ext}$, respectively. Arrows indicate a jump of the phase particle to an adjacent minimum, when the present local minimum disappears. 
(g) JPA resonant frequency $\omega_0$ when sweeping $\Phi_\mathrm{ext}$ towards larger (blue) and smaller (orange) values, calculated from Eq.~(\ref{eqn:WallquistWApprox}) with $L_\mathrm{loop}\,{=}\,0$, $L_\mathrm{r}\,{=}\,10L_\mathrm{s,min}$ and $L_\mathrm{s,min}\,{=}\,\Phi_0/(4 \pi I_\mathrm{c})$. Furthermore, the $\omega_0$ values of the corresponding position of the phase particle in panels (a-f) are denoted by green dots in panel (g).}
    \label{fig3}
\end{figure}

We now discuss hysteretic dc-SQUIDs for finite screening $\beta_\mathrm{L}\,{>}\,0$. To illustrate the behavior of a dc-SQUID in equilibrium, we consider a phase particle in the dc-SQUID potential~\cite{Lefevre:1992}
\begin{eqnarray}\label{eqn:PotSQUID}
\frac{U(\varphi_+,\varphi_-)}{E_\mathrm{J}}=&&2-2\cos\varphi_+\cos\varphi_-+2j_\mathrm{tr}\varphi_+\nonumber\\&&+\frac{2}{\pi\beta_\mathrm{L}}\left(\varphi_--\pi\frac{\Phi_\mathrm{ext}}{\Phi_0}\right)^2 \, ,
\end{eqnarray}
where $j_\mathrm{tr}\,{\equiv}\,I_\mathrm{tr}/2I_\mathrm{c}$ is the normalized transport current through the dc-SQUID and $E_\mathrm{J}\,{\equiv}\,I_\mathrm{c}\Phi_0/2\pi$ is the coupling energy of a single Josephson junction.  Figure~\ref{fig3} shows the potential $U(\varphi_+,\varphi_-)$ calculated according to Eq.~(\ref{eqn:PotSQUID}) for different values of the external flux $\Phi_\mathrm{ext}$ and with a typical experimental value of $\beta_\mathrm{L}\,{=}\,0.6$ as well as $j_\mathrm{tr}\,{=}\,0$. Note that for $j_\mathrm{tr}\,{=}\,0$ there is no tilt of the potential landscape along the $\varphi_+$-direction.
In contrast to other works investigating hysteretic dc-SQUIDs~\cite{Lefevre:1992,Balestro:2003,Hoskinson:2009}, in our experiments the transport current $j_\mathrm{tr}$ is negligible at all times, meaning that the possible equilibrium phase differences $(\varphi_+^\mathrm{min},\varphi_-^\mathrm{min})$ of the dc-SQUID are given by the local minima shown in Fig.~\ref{fig3}. 
One can consider a phase particle with coordinates $(\varphi_+^\mathrm{min},\varphi_-^\mathrm{min})$ residing in one of the minima. Due to the periodicity of the potential, multiple equivalent local minima which describe the same state of the dc-SQUID exist. For $\Phi_\mathrm{ext}\,{=}\,0$, all local minima are degenerate. However, for finite $\Phi_\mathrm{ext}$ and finite $\beta_\mathrm{L}$ there are two classes of minima corresponding to clock- and counter-clock-wise circulating screening currents. One is shifted upwards and the other one downwards in energy, therefore corresponding to a class of metastable and stable states, respectively. 

In order to obtain $\varphi_-^\mathrm{min}(\Phi_\mathrm{ext})$, one can track the equilibrium positions of the phase particle for a continuously varied external magnetic flux. Figure~\ref{fig3} illustrates the behavior of the phase particle (shown as a green dot) for increasing and decreasing external flux.
In the following, we neglect thermally activated and tunneling processes. That is, we assume that the phase particle resides in one specific minimum as long as this minimum exists. 
If the external flux $\Phi_\mathrm{ext}$ and therefore $\varphi_-^\mathrm{min}$ is changed such that this minimum vanishes, the phase particle will move to one of the two adjacent minima.
Because the adjacent minima always belong to a different class of local minima, one obtains a discontinuity in $\varphi_-^\mathrm{min}(\Phi_\mathrm{ext})$. The position of this discontinuity depends on the magnitude of $\beta_\mathrm{L}$. This discontinuity, in turn, leads to a discontinuity of the dc-SQUID inductance $L_\mathrm{s}(\Phi_\mathrm{ext})$, which can be experimentally observed as a jump of the JPA resonant frequency $\omega_0(\Phi_\mathrm{ext})$.
Evaluating Eq.~(\ref{eqn:WallquistWApprox}) with the obtained $\varphi_-^\mathrm{min}(\Phi_\mathrm{ext})$ can then be used to derive the flux dependence of the resonant frequency for the whole JPA circuit.

In addition, from the simulations we find that for finite $\beta_\mathrm{L}$ there is always a hysteretic behavior. However, for small $\beta_\mathrm{L}$ it only appears at frequencies much smaller than the maximum resonant frequency. Furthermore, a small hysteresis may also be smeared out by thermally activated switching between adjacent local energy minima.
In other words, the behavior observed in our devices is quite different from the textbook discussion of SQUID devices, where hysteretic behavior with respect to an external flux only occurs above a certain threshold value of the screening parameter.

The two-dimensional potential landscapes shown in Fig.~\ref{fig3} have been calculated for $j_\mathrm{tr}\,{=}\,0$. However, in our experiments shown below, a finite microwave signal is applied resulting in a finite microwave current across the dc-SQUID which resides in a current anti-node of the resonator. This finite microwave current results in a periodic tilt of the dc-SQUID potential along the $\varphi_+$-axis. Of course, this periodic tilt affects the dynamics of the phase particle. However, in our experiments the applied microwave signal is very small, resulting in microwave currents which are at least two orders of magnitude smaller than the junction critical current. As a consequence, the tiny periodic tilt of the dc-SQUID potential along the $\varphi_+$-axis can be neglected when analyzing the phase dynamics. Note that the finite microwave current results in a smearing of the hysteretic $\varphi_-^\mathrm{min}$ versus $\Phi_\mathrm{ext}$ dependence similar to thermal or other noise currents.

\section{Experimental measurements of the JPA flux dependence}\label{sec:ExpResults}

{\renewcommand{\arraystretch}{1.2}
\setlength{\tabcolsep}{5pt}
\begin{table*}
\caption{Parameters extracted from fitting of the flux-dependent JPA resonant frequency for different samples with the estimation of $L_\mathrm{r}\,{=}\,\unit[2]{nH}$. The external quality factors $Q_\mathrm{ext}$ and internal quality factors $Q_\mathrm{int}$ are obtained from independent fits of Eq.~(\ref{eqn:QualityFit}).}
\centering
\begin{ruledtabular}
\begin{tabular}{l*{8}{c}r}
Sample            & $I_\mathrm{c}$ ($\mu$A) & $\beta_\mathrm{L}$& $L_\mathrm{loop}$ (pH) & $\omega_\mathrm{r}/2\pi$ (GHz) & $E_\mathrm{J}/h$ ($\mathrm{THz}$) & $Q_\mathrm{ext}$ & $Q_\mathrm{int}$ \\
\hline
JPA~1	  & 2.45 & 0.09 & 35.8 & 5.808 & 1.22 & 300--360 & $>$30000 \\ 
JPA~2	  & 2.41 & 0.10 & 40.7 & 5.838 & 1.20 & 240--260 & $>$30000 \\ 
JPA~3	 & 11.39 & 0.55 & 49.9 & 6.220 & 5.66 & 5300 & 1200 \\ 
JPA~4	     & 9.82 & 0.55 & 57.2 & 6.164 & 4.88 & 11000 & 1100 \\ 
JPA~5	 & 9.64 & 0.56 & 52.9 & 6.216 & 4.79 & 72000 & 1300 \\ 
\end{tabular}
\end{ruledtabular}
\label{tab1}
\end{table*}
}

In order to experimentally investigate the properties of hysteretic JPAs, we study five samples (JPA~1 to JPA~5) with different screening parameters as well as different external and internal quality factors. While JPA~1 and JPA~2 have small screening parameters and high internal quality factors, JPA~3 to JPA~5 have larger screening parameters of $\beta_\mathrm{L}\simeq0.5$ and lower internal quality factors (see Tab.~\ref{tab1}). The larger $\beta_\mathrm{L}$ values for samples JPA~3 to JPA~5 are explained by higher critical currents $I_\mathrm{c}$ of the Josephson junctions while the loop inductance $L_\mathrm{loop}$ is similar for all samples. In the experiments discussed in this section, no additional external pump signal is applied to the JPA.

For resonators, the internal and external quality factors are important quantities. The internal quality factor $Q_\mathrm{int}$ provides information about the internal losses of the resonator while the external quality factor $Q_\mathrm{ext}$ is mainly given by the coupling capacitor $C_c$, which determines the coupling strength of the resonator to the signal port~\cite{Goetz:2016}. To extract the quality factors and the resonant frequencies for different external flux values, the expression~\cite{Yamamoto:2016}
\begin{equation}\label{eqn:QualityFit}
S_{11}=\frac{(\omega-\omega_0)^2\,{+}\,i\kappa_\mathrm{int}(\omega-\omega_0)\,{+}\,(\kappa_\mathrm{ext}^2-\kappa_\mathrm{int}^2)/4}{\left[(\omega-\omega_0)\,{+}\,i(\kappa_\mathrm{ext}\,{+}\,\kappa_\mathrm{int})/2\right]^2}
\end{equation}
for the reflection coefficient, obtained from an input-output theory, is fitted to the experimental data. Here, $\kappa_\mathrm{ext}\,{=}\,\omega_0/Q_\mathrm{ext}$ and $\kappa_\mathrm{int}\,{=}\,\omega_0/Q_\mathrm{int}$ are the external and the internal loss rates, respectively.
The measurements are performed in the low power regime with less than one photon on average inside the resonator.
For samples JPA~1 and JPA~2, we extract internal quality factors of up to $Q_\mathrm{int}\,{=}\,3\times10^4$ depending on the external flux. Both samples show an overcoupled behavior. Furthermore, we observe a strongly undercoupled behavior for samples JPA~3 to JPA~5, where the external quality factor is much larger than the internal quality factor. Moreover, the internal quality factors are lower than those of samples JPA~1 and JPA~2.

\begin{figure}
  \begin{center}
    \includegraphics{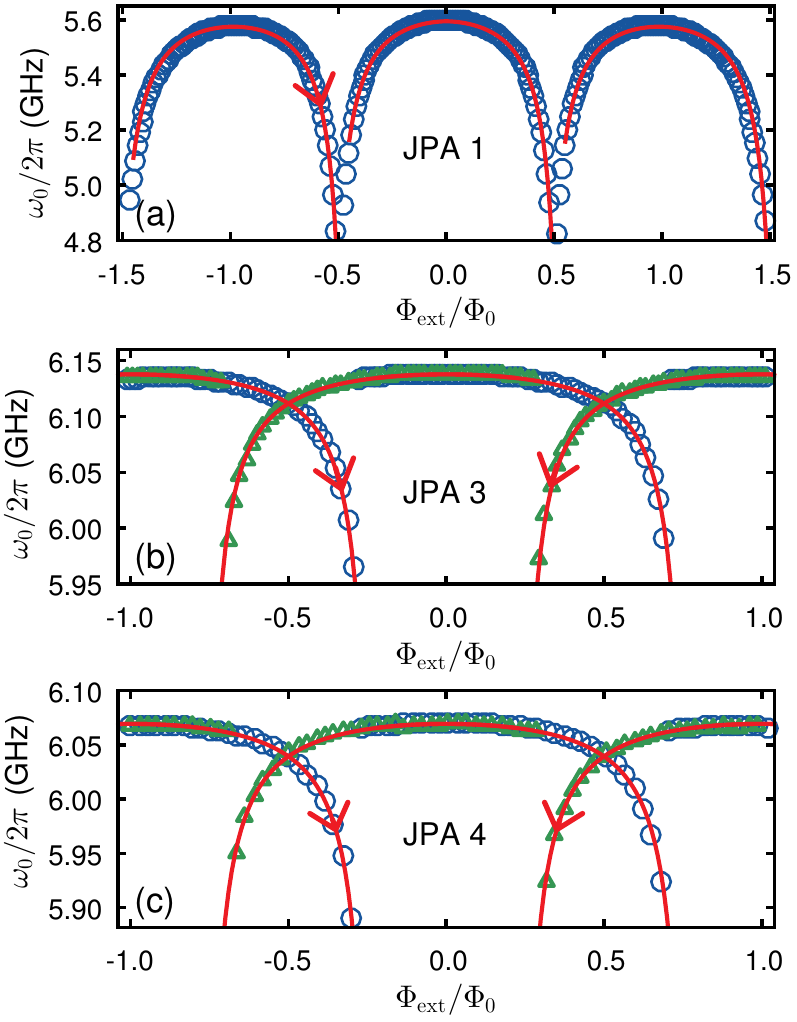}
  \end{center}
  \caption{JPA Resonant frequency $\omega_0$ of \textbf{(a)} JPA~1, \textbf{(b)} JPA~3 and \textbf{(c)} JPA~4 as a function of the applied flux $\Phi_\mathrm{ext}$ as well as numerical fits (red lines). Blue circles and green triangles mark the data taken for increasing and decreasing $\Phi_\mathrm{ext}$, respectively, with arrows further indicating the sweep direction. The JPAs are stabilized at temperatures between \unit[17]{mK} and \unit[50]{mK}.
The fitting results are summarized in Tab.~\ref{tab1}.}
\label{fig4}
\end{figure}

Figure~\ref{fig4} shows the flux-dependent JPA resonant frequencies for three samples together with numerical fits according to Eq.~(\ref{eqn:WallquistWApprox}). In the experimentally accessed frequency range, Eq.~(\ref{eqn:WallquistWApprox}) holds with an error of less than $0.5\%$ compared to the exact solution from Eq.~(\ref{eqn:WallquistW}).
Using $L_\mathrm{r}\,{=}\,\unit[2]{nH}$ estimated from the geometric design parameters and the known characteristic impedance $Z_0\,{=}\,\unit[50]{\Omega}$, one can extract several relevant JPA parameters from the data as summarized in Tab.~\ref{tab1}. 
In order to fit the flux dependence of JPA~1 properly, an additional effect leading to a decrease of the maximal JPA resonant frequency with increasing $\Phi_\mathrm{ext}$ has to be included. We use a simple linear model which decreases the junction critical current $I_\mathrm{c}$ proportional to $\Phi_\mathrm{ext}$. The effect may originate from a parasitic in-plane component of the magnetic field generated by the superconducting coil which is penetrating the insulating tunnel barrier of the Josephson junctions.

For the two samples, JPA~1 and JPA~2, no hysteretic behavior is experimentally observed [see Fig.~\ref{fig4}(a)]. All in all, the fit describes the experimental data very well. However, there are deviations between data and theory predictions at the left part of each period where the simulation predicts a slightly hysteretic behavior which is not reproduced by the experimental data. We attribute the observed deviation to a finite noise floor which is not included in the simulation and which causes a premature hopping of the phase particle to an adjacent minimum. Therefore, it is expected that the hysteresis is not as pronounced in the experimental data as predicted by simulations.

For the three other samples, JPA~3 to JPA~5, a strong hysteretic behavior is observed in the resonant frequency versus applied flux dependence. Figure~\ref{fig4}(b) and Figure~\ref{fig4}(c) show an overlay of both sweep directions of the external flux. The flux dependence is described very well by the model calculations. The hysteresis over a large frequency window is explained by larger screening parameters $\beta_\mathrm{L}$ as compared to JPA~1 and JPA~2. With increasing $\beta_\mathrm{L}$, the rigid coupling between the two phase differences across the junctions is lost, allowing for multiple classes of minimal energy states of the dc-SQUID for a given external flux. These different classes of minimal energy states manifest themselves in different resonant frequencies of the JPA.
For the measured samples, we find only two such classes whereas even more can exist for sufficiently large $\beta_\mathrm{L}$. Depending on the history of the dc-SQUID regarding the external flux, different eligible minimal energy states are occupied by the dc-SQUID and, thus, a hysteretic behavior when changing the sweep direction of the external flux is observed.

\section{Nondegenerate gain}\label{sec:GainMeasurements}
\begin{figure}
        \begin{center}
        \includegraphics{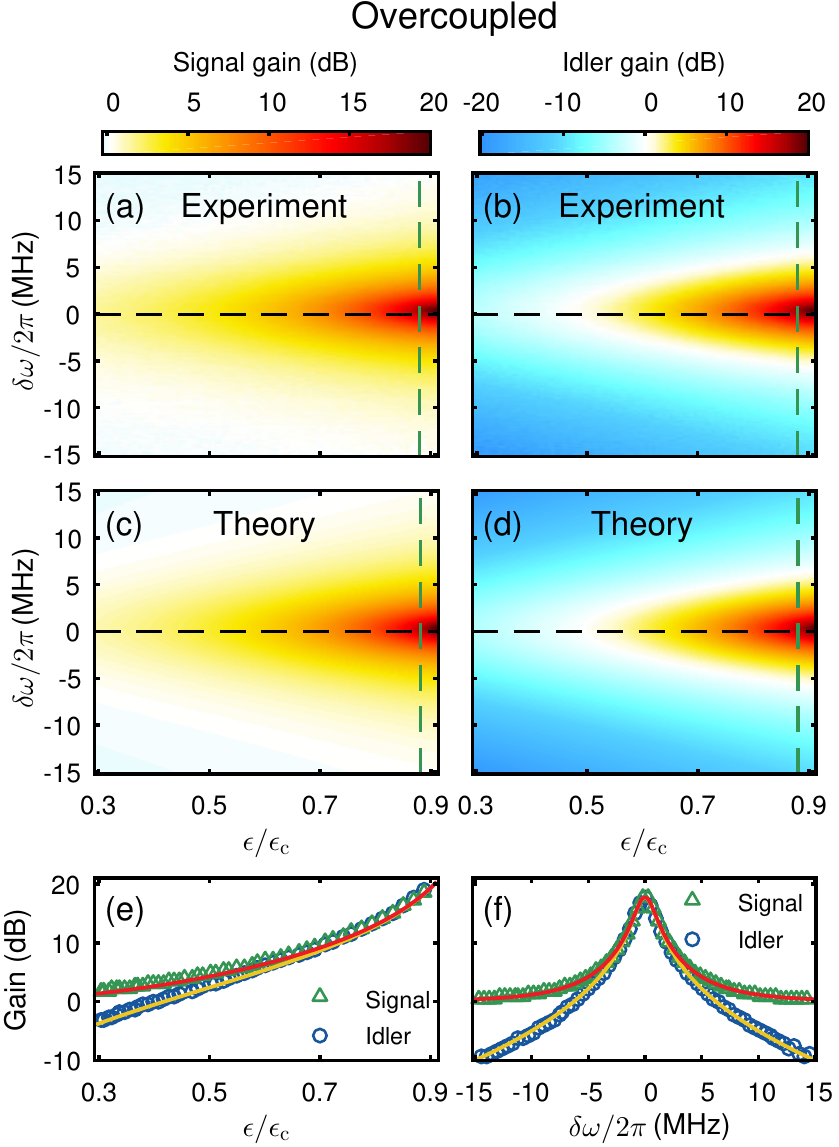}
        \end{center}
    \caption{(a,b) Experimental spectra of the nondegenerate signal and idler gain as a function of $\epsilon$ and signal detuning $\delta \omega$ from half the pump frequency of $\omega_\mathrm{pump}/4\pi\,{=}\,\unit[5.4]{GHz}$ for JPA~1.
    (c,d) Theoretical calculations with $g\simeq\unit[3.49]{V^{-1}}$ of the signal and idler gain computed from Eqs. (\ref{eqn:YamamotoTheorySig}) and (\ref{eqn:YamamotoTheoryId}), respectively.
    (e) Signal and idler gain as a function of $\epsilon$ along the black dashed lines.
    (f) Signal and idler gain as a function of the signal frequency along the green dashed lines. The symbols mark the experimental data and solid lines are fits of the data by Eqs. (\ref{eqn:YamamotoTheorySig}) and (\ref{eqn:YamamotoTheoryId}). The JPA temperature is stabilized at $\unit[50]{mK}$.}
    \label{fig5}
\end{figure}

\begin{figure}
        \begin{center}
        \includegraphics{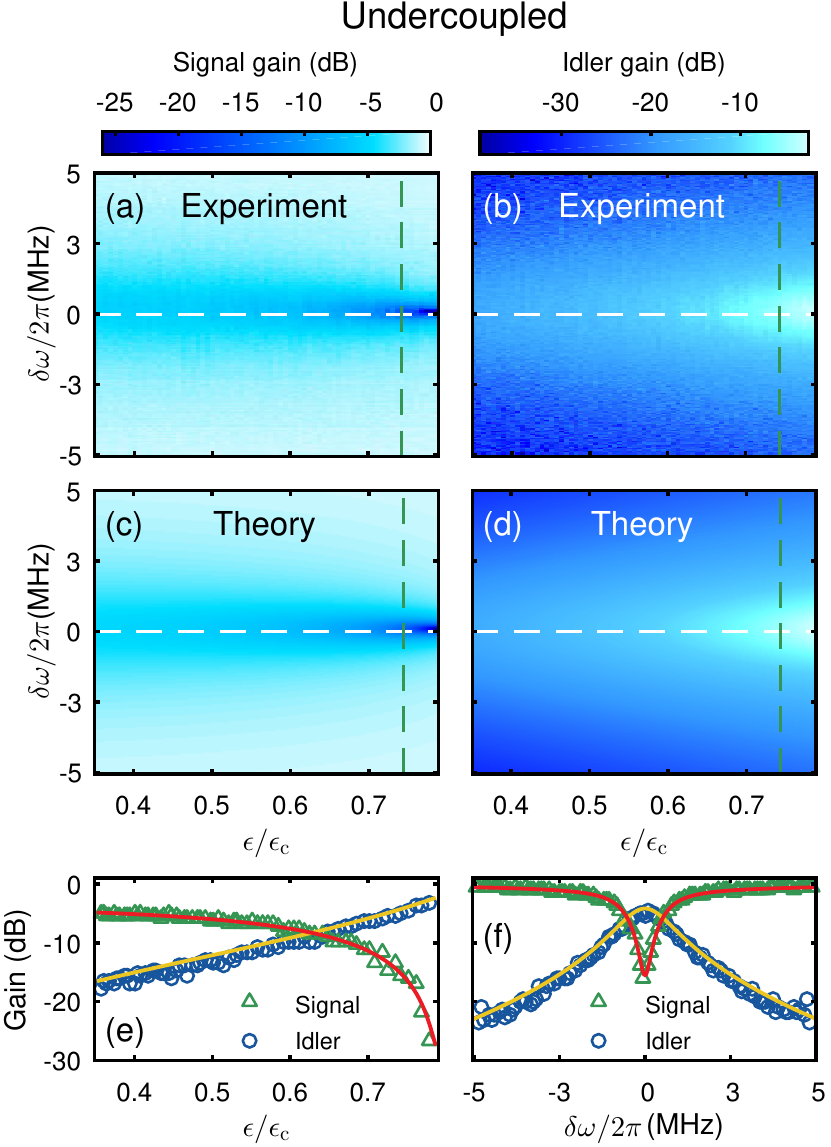}
        \end{center}
    \caption{(a,b) Experimental spectra of the nondegenerate signal and idler gain as a function of $\epsilon$ and signal detuning $\delta \omega$ from half the pump frequency of $\omega_\mathrm{pump}/4\pi\,{=}\,\unit[6.125]{GHz}$ for JPA~3.
    (c,d) Theoretical calculations with $g\simeq\unit[0.17]{V^{-1}}$ of the signal and idler gain computed from Eqs. (\ref{eqn:YamamotoTheorySig}) and (\ref{eqn:YamamotoTheoryId}), respectively.
    (e) Signal and idler gain as a function of $\epsilon$ along the white dashed lines.
    (f) Signal and idler gain as a function of the signal frequency along the green dashed lines. The symbols mark the experimental data and solid lines are fits of the data by Eqs. (\ref{eqn:YamamotoTheorySig}) and (\ref{eqn:YamamotoTheoryId}). The JPA temperature is stabilized at $\unit[30]{mK}$.}
    \label{fig6}
\end{figure}

In this section, we investigate the nondegenerate gain of two JPAs, where one JPA has a non-hysteretic and overcoupled behavior and the other one has a hysteretic and undercoupled behavior. To this end, a flux value corresponding to a certain resonant frequency $\omega_0$ of the JPA is fixed. Then, a pump tone with frequency $\omega_\mathrm{pump}\,{=}\,2\omega_0$ is applied to the JPA. 
Regarding the input signal, the JPA is operated in the nondegenerate mode, meaning that the frequency of the applied signal $\omega_\mathrm{s}\,{=}\,\omega_\mathrm{pump}/2\,{+}\,\delta \omega$ always has a finite offset from half the pump frequency, $\delta \omega\,{\neq}\,0$. In order to evaluate the experimental data we use explicit expressions for the nondegenerate gain for the case of a flux-driven JPA. The nondegenerate signal gain $G_\mathrm{s}(\delta \omega)$ and idler gain $G_\mathrm{i}(\delta \omega)$ for $\omega_\mathrm{pump}\,{=}\,2\omega_0$ are given by~\cite{Yamamoto:2016}
\begin{equation}\label{eqn:YamamotoTheorySig} 
G_\mathrm{s}(\delta \omega)=\frac{\kappa_\mathrm{int}^2\delta\omega^2+\left[(\kappa_\mathrm{int}^2-\kappa_\mathrm{ext}^2)/4-\epsilon^2\omega_0^2-\delta\omega^2\right]^2}{\kappa_\mathrm{tot}^2\delta\omega^2+\left[\kappa_\mathrm{tot}^2/4-\epsilon^2\omega_0^2-\delta\omega^2\right]^2}\, ,
\end{equation}
\begin{equation}\label{eqn:YamamotoTheoryId} 
G_\mathrm{i}(\delta \omega)=\frac{\kappa_\mathrm{ext}^2\epsilon^2\omega_0^2}{\kappa_\mathrm{tot}^2\delta\omega^2+\left[\kappa_\mathrm{tot}^2/4-\epsilon^2\omega_0^2-\delta\omega^2\right]^2}\, ,
\end{equation}
where $\kappa_\mathrm{tot}$ is the total resonator loss and $\epsilon\,{=}\,gA_\mathrm{pump}$ is related to the root-mean-squared pump amplitude $A_\mathrm{pump}$ at the sample box via a coupling constant $g$. Eqs.~(\ref{eqn:YamamotoTheorySig}) and (\ref{eqn:YamamotoTheoryId}) are only valid for $\epsilon\leq\epsilon_\mathrm{c}\,{=}\,\kappa_\mathrm{tot}/2 \omega_0$. 

Figure~\ref{fig5}(a) and Figure~\ref{fig5}(b) show the nondegenerate signal and idler gain as a function of $\epsilon$ for JPA~1 (non-hysteretic, overcoupled). The pump frequency is fixed at $\omega_\mathrm{pump}/2\pi\,{=}\,\unit[10.8]{GHz}$ corresponding to a flux working point of $\Phi_\mathrm{ext}\,{=}\,0.39\,\Phi_0$. The idler gain is measured by sweeping a signal at frequency $\omega_\mathrm{s}\,{=}\,\omega_\mathrm{pump}/2\,{+}\,\delta \omega$ and comparing to the generated idler mode at frequency $\omega_\mathrm{i}\,{=}\,\omega_\mathrm{pump}/2\,{-}\,\delta \omega$. Amplification can only be observed within a frequency window defined by the resonator bandwidth and centered at the resonant frequency. In this region one observes an increased gain for both the signal and idler mode with increasing pump power. Theoretical predictions from Eqs.~(\ref{eqn:YamamotoTheorySig}) and (\ref{eqn:YamamotoTheoryId}) are depicted in Fig.~\ref{fig5}(c) and Fig.~\ref{fig5}(d) for the signal and the idler mode, respectively. Only the coupling constant $g\simeq\unit[3.49]{V^{-1}}$ is used as a fitting parameter, while the quality factors and the resonant frequency are fixed to previously determined experimental values. The given values for $g$ include an additional uncertainty due to the fact that $A_\mathrm{pump}$ is calculated from the power set at the microwave source using an estimated pump line attenuation of $\unit[61]{dB}$.
Figure~\ref{fig5}(e) and Figure~\ref{fig5}(f) show corresponding cuts through the experimental and theoretical spectra. As it can be seen, the model reproduces both the signal and idler modes very well.

In addition, sample JPA~3 (hysteretic, undercoupled) is investigated in the nondegenerate mode by applying a fixed pump tone of frequency $\omega_\mathrm{pump}/2\pi\,{=}\,\unit[12.25]{GHz}$. The flux working point is $\Phi_\mathrm{ext}\,{=}\,{-}\,0.40\,\Phi_0$.
The experimentally obtained spectra of the signal and idler as a function of the pump power are depicted in Fig.~\ref{fig6}(a) and Fig.~\ref{fig6}(b). For $\delta \omega/2\pi\,{\simeq}\,\unit[10]{kHz}$, the incident signal is increasingly deamplified by up to $\unit[-30]{dB}$ with increasing pump power, while the idler gain increases. Since these undercoupled devices are described well by our parametric amplifier theory, we still call them JPAs although they do not act as amplifiers but rather as attenuators. Again Fig.~\ref{fig6}(c) and Fig.~\ref{fig6}(d) depict theoretical predictions with $g\simeq\unit[0.17]{V^{-1}}$ as the only fitting parameter. They reproduce the experimentally observed behavior accurately.
The deamplification behavior of JPA~3 is in strong contrast to sample JPA~1 where the signal gain increases with increasing pump power. To understand this behavior qualitatively, we simplify Eq.~(\ref{eqn:YamamotoTheorySig}) for $\delta \omega\,{\rightarrow}\,0$ and obtain
\begin{equation}\label{eqn:YamamotoTheorySigApprox} 
G_\mathrm{s}\approx\frac{\left[(\kappa_\mathrm{int}^2-\kappa_\mathrm{ext}^2)/4-\epsilon^2\omega_0^2\right]^2}{\left[\kappa_\mathrm{tot}^2/4-\epsilon^2\omega_0^2\right]^2}\, .
\end{equation}
If the sample is overcoupled ($\kappa_\mathrm{ext}\,{>}\,\kappa_\mathrm{int}$), the numerator of Eq.~(\ref{eqn:YamamotoTheorySigApprox}) is monotonously increasing while the denominator is monotonously decreasing with increasing $\epsilon$. However, for an undercoupled JPA ($\kappa_\mathrm{ext}\,{<}\,\kappa_\mathrm{int}$) the numerator crosses zero for a certain threshold value $\epsilon\,{=}\,\epsilon_\mathrm{crit}{\,<\,}\epsilon_\mathrm{c}$, meaning that the signal is increasingly deamplified with increasing $\epsilon$ until the threshold is reached.
In the overcoupled regime, the JPA acts as an amplifier for an incident signal, whereas for undercoupled JPAs the incident signal is deamplified depending on the pump power. In this case, the device is no longer an amplifier but acts as a tunable microwave attenuator. 

\section{Conclusion}\label{sec:Summary}
In summary, we have developed an efficient approach to describe both the hysteretic and non-hysteretic dependence of the resonant frequency $\omega_0$ of flux-driven Josephson parametric amplifiers on the applied magnetic flux. We have achieved this by classical simulations of the position of the phase particle in the flux-dependent two-dimensional potential landscape of the dc-SQUID which strongly depends on the screening parameter $\beta_\mathrm{L}$. Furthermore, we have applied a distributed-element model to describe $\omega_0$ of the JPA.

We conclude that even for small screening parameters $\beta_\mathrm{L}\,{\ll}\,1$ one obtains a quantitatively different dependence of $\omega_0$ on the external flux as compared to the case with $\beta_\mathrm{L}\,{=}\,0$. The observed hysteretic and non-hysteretic behavior of the investigated JPAs is clearly related to different magnitudes of their screening parameters $\beta_\mathrm{L}$ arising from different critical current values of their Josephson junctions. The JPA resonant frequency thus provides a non-invasive readout of the dc-SQUID state which can be used to prepare the dc-SQUID in a specific state. Additionally, the nondegenerate gains of JPAs with different resonator characteristics have been investigated. We have observed that, in the undercoupled regime, a JPA can act as a tunable microwave attenuator with attenuation of up to $\unit[-30]{dB}$, while an overcoupled JPA acts as a linear amplifier, as expected. The experimentally observed behavior is accurately reproduced by model calculations based on a simple, linearized and explicit formalism for the flux-driven JPA.

\section*{Acknowledgments}
We acknowledge support by the German Research Foundation through SFB 631 and FE 1564/1-1, the EU projects PROMISCE and SCALEQIT, the Elite Network of Bavaria through the program ExQM, the International Max Planck Research School "Quantum Science and Technology", JSPS KAKENHI (Grant No. 26220601, 15K17731) as well as the ImPACT Program of Council for Science. We would like to thank K. Kusuyama for assistance with part of the JPA fabrication.

\end{document}